\documentstyle[aps,epsf,floats,amsfonts,amssymb,amsmath]{revtex}
\newcommand{\bea}{\begin{eqnarray}}\newcommand{\eea}{\end{eqnarray}}
\newcommand{\brr}{\begin{array}}\newcommand{\err}{\end{array}}
\newcommand{\bit}{\begin{itemize}}\newcommand{\eit}{\end{itemize}}
\newcommand{\ben}{\begin{enumerate}}\newcommand{\een}{\end{enumerate}}

\def\lan{\langle}
\def\lf{\left}

\def\ran{\rangle}

\def\ri{\right}

\def\al{\alpha}\def\bt{\beta}

\def\te{\theta}

\def\si{\sigma}
\def\om{\omega}

\def\1{{_{1}}}
\def\2{{_{2}}}
\begin{document}

\title{Neutrino mixing contribution to the cosmological constant}
\author{M.~Blasone${}^{\flat\sharp}$, A.~Capolupo${}^{\flat}$,
S.~Capozziello${}^{\flat}$, S.~Carloni${}^{\flat}$ and
G.~Vitiello${}^{\flat \sharp }$ \vspace{3mm}}

\address{ ${}^{\flat}$ Dipartimento di  Fisica "E.R. Caianiello" and INFN,
Universit\`a di Salerno, I-84100 Salerno, Italy
\\ [2mm] ${}^{\sharp}$ Unit\`a INFM, Salerno, Italy
\vspace{2mm}}

\date{{\bf Version}, \today}

\maketitle

\begin{abstract}

We show that the non--perturbative vacuum structure associated
with neutrino mixing leads to a non--zero contribution to the
value of the cosmological constant. Such a contribution comes from
the specific nature of the mixing phenomenon. Its origin is
completely different from the one of the ordinary contribution of
a massive spinor field. We estimate this neutrino mixing
contribution by using the natural cut--off appearing in the
quantum field theory formalism for neutrino mixing and
oscillation.

\end{abstract}

\vspace{8mm}

\section{Introduction}

By resorting to the recent discovery of the unitary inequivalence
between the mass and the flavor vacua for neutrino fields in
quantum field theory (QFT)
\cite{BV95,hannabuss,Ji,fujii,remarks,remarks2,bosonmix,comment},
we show that the non--perturbative vacuum structure associated
with neutrino mixing may lead to a non--zero contribution to the
value of the cosmological constant.

The contribution we find comes from the specific nature of the
field mixing and is therefore of different origin with respect to
the ordinary well known vacuum energy contribution of a massive
spinor field.

The nature of the cosmological constant, say $\Lambda$, is one of
the most intriguing issues in  modern theoretical physics and
cosmology. Data coming from observations indicate that not only
$\Lambda$ is different from zero, but it also dominates the
universe dynamics driving an accelerated expansion (see for
example \cite{WMAP,sahni}).

In the classical framework, $\Lambda$ can be considered as an
intrinsic (i.e. not induced by matter) curvature of  space-time or
a sort of shift in the matter Lagrangian. In the latter case
$\Lambda$ can be considered as an unclustered, non interacting
component of the cosmic fluid with a constant energy density
$\rho$ and the equation of state $p=-\rho$. These properties and
the fact that the presence of a cosmological constant fluid has to
be compatible with the structure formation, allow to set the upper
bound $\Lambda < 10^{-56} cm^{-2}$ \cite{Zeldovic1}.

In the quantum framework, the standard approach is to consider the
cosmological constant as a gravitational effect of vacuum energy
\cite{sahni,Zeldovic2,weinberg,Gell}. A common problem of all
these approaches is that they do not provide a value of $\Lambda$
in the bound given above. This is known as the {\it cosmological
constant problem}\cite{sahni} and it has been tried to address it
in different ways (see for example \cite{Cap}).

In this paper, we show  that the vacuum energy induced by  the
neutrino mixing may contribute to the value of cosmological
constant in a  fundamentally different way from the usual
zero-point energy contribution, as already stated above.

Indeed,  it has been realized \cite{BV95,remarks} that the mixing
of massive neutrino fields is a highly non-trivial transformation
in QFT. The vacuum for neutrinos with definite mass is not
invariant under the mixing transformation and in the infinite
volume limit it is unitarily inequivalent to the vacuum for the
neutrino fields with definite flavor number. This affects the
oscillation formula which turns out to be different from the usual
Pontecorvo formula \cite{Pontec} and a number of consequences have
been discussed \cite{remarks}.

The existence of the two inequivalent vacua for the flavor and the
mass eigenstate neutrino fields, respectively, is crucial in order
to obtain a non--zero contribution to the cosmological constant as
we show below.

In Section II, we shortly summarize the QFT formalism for neutrino
mixing. In Section III, we show that the neutrino contribution to
the value of the cosmological constant is non--zero and then we
estimate its value by using the natural scale of neutrino mixing
as cut--off. The result turns out to be compatible with the above
mentioned upper bound on $\Lambda$. Section IV is devoted to the
conclusions. The appendix is devoted to the tetradic formalism in
the Friedmann Robertson Walker (FRW) space--time.

\section{Neutrino mixing in Quantum Field Theory}

The main features of the QFT formalism for the neutrino mixing are
summarized below. For simplicity we restict ourselves to the two
flavor case. Extension to three flavors can be found in refs.
\cite{comment}. The following Lagrangian density describes the
Dirac neutrino fields with a mixed mass term:

\bea\label{lagemu}{\cal L}(x)\,= \,{\bar \Psi_f}(x) \lf( i
\not\!\partial-M \ri) \Psi_f(x)\  . \eea

The relation between Dirac fields $\Psi_f(x)$, eigenstates of
flavor, and Dirac fields $\Psi_m(x)$, eigenstates of mass, is
given by

\bea\label{fermix} \Psi_f(x) \, = {\cal U} \, \Psi_m (x). \eea
${\cal U} $ is the mixing matrix and $\Psi_m^T=(\nu_1,\nu_2)$. The
mixing matrix is

\bea \label{fermix1} {\cal U}=\begin{pmatrix}
  \cos\theta & \sin\theta\\
  -\sin\theta & \cos\theta
\end{pmatrix}
\eea being $\te$ the mixing angle. Using Eq.(\ref{fermix1}), we
diagonalize the quadratic form of Eq.(\ref{lagemu}), which then
reduces to the Lagrangian for the Dirac fields $\Psi_m(x)$, with
masses $m_i$, $ i =1,2 $ :

\bea\label{lag12} {\cal L}(x)\,=\,  {\bar \Psi_m}(x) \lf( i
\not\!\partial -  \textsf{M}_d\ri) \Psi_m(x)  \, , \eea where
$\textsf{M}_d = diag(m_1,m_2)$. The mixing transformation
(\ref{fermix}) can be written as \cite{BV95}

\bea\label{mt} &&\nu_{\si}(x)\equiv G^{-1}_{\bf \te}(t) \,
\nu_{i}(x)\, G_{\bf \te}(t), \eea where $(\si,i)=(e,1), (\mu,2)$,
and the generator $G_{\bf \te}(t)$ is given by

\bea G_{\bf \te}(t)=\exp\Big[\te\int d^{3}{\bf
x}\lf(\nu_{1}^{\dag}(x)\nu_{2}(x)-\nu_{2}^{\dag}(x)\nu_{1}(x)\ri)
\Big]  . \eea

The free fields  $\nu_i$ (i=1,2) are given, in the usual way, in
terms of creation and annihilation operators (we use $t\equiv
x_0$):

\bea\label{2.2} \nu_{i}(x) = \sum_{r} \int \frac{d^3{\bf
k}}{(2\pi)^\frac{3}{2}} \lf[u^{r}_{{\bf k},i}(t) \al^{r}_{{\bf
k},i}\:+    v^{r}_{-{\bf k},i}(t) \bt^{r\dag }_{-{\bf k},i}  \ri]
e^{i {\bf k}\cdot{\bf x}} ,\qquad i=1,2 \, \eea with $u^{r}_{{\bf
k},i}(t)=e^{-i\om_{k,i} t}u^{r}_{{\bf k},i}$, ~ $v^{r}_{{\bf
k},i}(t)=e^{i\om_{k,i} t}v^{r}_{{\bf k},i}$ ~ and
$\om_{k,i}=\sqrt{{\bf k}^2+m_i^2}$. The mass eigenstate vacuum is
denoted by $|0\ran_{m}$:  $\; \; \al^{r}_{{\bf k},i}|0\ran_{m}=
\bt^{r }_{{\bf k},i}|0\ran_{m}=0$. The anticommutation relations,
the wave function orthonormality and completeness relations are
the usual ones (cf. Ref. \cite{BV95}).

The flavor fields are obtained from Eq. (\ref{mt}):

\bea\label{exnue1} &&{}\quad\qquad \nu_\si(x)= \sum_{r} \int
\frac{d^3{\bf k}}{(2\pi)^\frac{3}{2}} \lf[ u^{r}_{{\bf k},i}(t)
\al^{r}_{{\bf k},\si}(t) + v^{r}_{-{\bf k},i}(t)
\bt^{r\dag}_{-{\bf k},\si}(t) \ri]  e^{i {\bf k}\cdot{\bf x}}
,\quad (\si,i)=(e,1) , (\mu,2) . \eea

The flavor annihilation operators are defined as $\al^{r}_{{\bf
k},\si}(t) \equiv G^{-1}_{\bf \te}(t)\al^{r}_{{\bf k},i} G_{\bf
\te}(t)$ and $\bt^{r}_{{-\bf k},\si}(t)\equiv
 G^{-1}_{\bf \te}(t) \bt^{r\dag}_{{-\bf k},i}
G_{\bf \te}(t)$. They annihilate the flavor vacuum $|0(t)\ran_{f}$
given by

\bea\label{flavac}
|0(t)\ran_{f}\,\equiv\,G_{\te}^{-1}(t)\;|0\ran_{m} \;. \eea

The crucial remark is that, in the infinite volume limit, the
vacuum $|0(t)\ran_{f}$ for the flavor fields and the vacuum
$|0\ran_{m}$ for the fields with definite masses are unitarily
inequivalent vacua \cite{BV95}.  For a rigorous general proof of
such inequivalence for any number of generations see
Ref.\cite{hannabuss}.

In Ref.\cite{remarks2} it was shown that the correct vacuum to be
used in the calculation of oscillation formulas is the flavor
vacuum $|0 {\rangle}_{f}$. Indeed, using the mass vacuum $|0
{\rangle}_{m}$ leads to violation of probability conservation
\cite{remarks2}. We will thus use $|0 {\rangle}_{f}$ in our
computations in the following.

For further reference, it is useful to list the explicit
expressions for the flavor annihilation/creation operators. In the
reference frame ${\bf k}=(0,0,|{\bf k}|)$ they are given by
\cite{BV95}:

\bea \al_{{\bf k},e}^{r}(t)&=&\cos\theta\;\al_{{\bf k},1}^{r} +
\sin\theta\lf(U_{\bf k}^*(t)\;\al_{{\bf k},2}^{r} +\epsilon^{r}
V_{\bf k}(t)\;\bt_{-{\bf k},2}^{r\dag}\ri) \;,
\\[2mm]
\al_{{\bf k},\mu}^{r}(t)&=& \cos\theta\;\al_{{\bf k},2}^{r} -
\sin\theta\lf(U_{\bf k}(t)\;\al_{{\bf k},1}^{r} - \epsilon^{r}
V_{\bf k}(t)\;\bt_{-{\bf k},1}^{r\dag}\ri) \;,
\\[2mm]
\bt^{r}_{-{\bf k},e}(t)&=&\cos\theta\;\bt_{-{\bf k},1}^{r} + \sin
\theta \lf(U_{\bf k}^*(t)\;\bt_{-{\bf k},2}^{r}
-\epsilon^{r}V_{\bf k}(t)\;\al_{{\bf k},2}^{r\dag}\ri) \;,
\\[2mm]
\bt^{r}_{-{\bf k},\mu}(t)&=&\cos\theta\;\bt_{-{\bf k},2}^{r} -\sin
\theta \lf(U_{\bf k}(t)\;\bt_{-{\bf k},1}^{r} +\epsilon^{r}V_{\bf
k}(t)\;\al_{{\bf k},1}^{r\dag}\ri) \;. \eea

$U_{\bf k}$ and $V_{\bf k}$  are  Bogoliubov coefficients given
by:

\bea V_{\bf k}(t)=|V_{\bf
k}|\;e^{i(\om_{k,2}+\om_{k,1})t}\;\;\;\;,\;\;\; \; U_{\bf
k}(t)=|U_{\bf k}|\;e^{i(\om_{k,2}-\om_{k,1})t} , \eea

\bea &&|U_{\bf k}|=\lf(\frac{\om_{k,1}+m_{1}}{2\om_{k,1}}\ri)
^{\frac{1}{2}}
\lf(\frac{\om_{k,2}+m_{2}}{2\om_{k,2}}\ri)^{\frac{1}{2}}
\lf(1+\frac{|{\bf k}|^{2}}{(\om_{k,1}+m_{1})
(\om_{k,2}+m_{2})}\ri) \equiv \cos(\xi^{{\bf k}}) ,
\\ \label{V}
&&|V_{\bf k}|=\lf(\frac{\om_{k,1}+m_{1}}{2\om_{k,1}}\ri)
^{\frac{1}{2}}
\lf(\frac{\om_{k,2}+m_{2}}{2\om_{k,2}}\ri)^{\frac{1}{2}}
\lf(\frac{|{\bf k}|}{(\om_{k,2}+m_{2})}-\frac{|{\bf
k}|}{(\om_{k,1}+m_{1})}\ri) \equiv \sin(\xi^{{\bf k}}) , \eea

\bea |U_{\bf k}|^{2}+|V_{\bf k}|^{2}=1. \eea

The function $|V_{\bf k}|$ is related to the condensate content of
the flavor vacuum\cite{BV95}: it is indeed zero for $m_{1} =
m_{2}$, it has a maximum at $|{\bf k}|=\sqrt{m_\1 m_\2}$ and, for
$ |{\bf k}| \gg\sqrt{m_\1 m_\2}$, it goes like $|V_{{\bf
k}}|^2\simeq (m_\2 -m_\1)^2/(4 |{\bf k}|^2)$.

\section{neutrino mixing contribution to the cosmological constant}

The connection between the vacuum energy density
$\lan\rho_{vac}\ran$ and the cosmological constant $\Lambda$ is
provided by the well known relation
\bea\label{ed} \lan\rho_{vac}\ran= \frac{\Lambda}{4\pi G} , \eea
where $G$ is the gravitational constant.

In order to calculate $\lan\rho_{vac}\ran$ we have to consider
tetrads and  spinorial connection. The symmetries of the
cosmological metric make this task easier.

Tetrads are defined as a local inertial coordinate system at every
space-time point defined by $g_{\mu
\nu}=e^{a}_{\mu}e^{b}_{\nu}\eta_{a b}$, where $g_{\mu \nu}$ it the
curved space-time metric and $\eta_{a b}$ is the Minkowski metric.
In the tetrads framework the parallel transport in a torsion free
space-time is defined by the affine spin connection one form $
\omega^{a}_{b}=\omega^{a}_{b \mu}d x^{\mu},$ which satisfies the
Cartan structure equations
\begin{equation}\label{eq forma di connessione}
  d e^{a} + \omega^{a}_{b}\wedge e^{b}=0,
\end{equation}
where $e^{a}$ is the tetrad one-form defined by
$e^{a}=e^{a}_{\nu}dx^{\nu}$. The energy--momentum tensor $T_{\mu
\nu}$ is obtained  by varying the action with respect to the
metric $g_{\mu\nu}$:

 \bea\ \label{impulsoenergia}
T_{\mu\nu}=\frac{2}{\sqrt{-g}}\frac{\delta S}{\delta
g^{\mu\nu}(x)} ,
 \eea
where the action is
 \bea\
S=\int \sqrt{-g}{\cal L}(x) d^{4}x.
 \eea%

In the present case, the energy momentum tensor is given by

\bea\label{tmn}
 T_{\mu\nu}(x) = \frac{i}{2}\left({\bar
\Psi}_{m}(x)\gamma_{\mu} \overleftrightarrow{D}_{\nu}
\Psi_{m}(x)\right)\ ,
 \eea
where $\overleftrightarrow{D}_{\nu}$ is the covariant derivative:
\begin{equation}\label{der spin}
D_{\nu}=\partial_{\nu}+\Gamma_{\nu},
 \quad\;\quad\;\quad  \Gamma_{\nu}=\frac{1}{8} \omega^{a
b}_{\nu}[\gamma_{a},\gamma_{b}], \quad\;\quad\;\quad
  \gamma_{\mu}(x)=\gamma^{c}e_{c \mu}(x),
\end{equation}
 being $\gamma^{c}$ the standard Dirac matrices, and
 $\bar{\Psi}\overleftrightarrow{D}_{\nu}\Psi=\bar{\Psi}D\Psi-(D\bar{\Psi})
\Psi$. Let us consider the FRW metric of the form
\begin{equation}\label{metrica di friedmann}
  ds^{2}=dt^{2}-a^{2}(t)\left(\frac{dr^{2}}{\rho}+r^{2}d\theta^{2}+r^{2}\sin^{2}
(\theta)d\phi^{2}\right),
\end{equation}
with $\rho=1-\kappa r^{2}$, and $\kappa=0,1,-1$. In the metric
(\ref{metrica di friedmann}), the most natural choice for tetrads
is
\begin{equation}\label{tetradi di Friedmann}
e^{0}=e^{0}_{0}d x^{0}= dt,\qquad e^{1}=e^{1}_{1}d
x^{1}=\frac{a(t)}{\rho}d r,\qquad e^{2}=e^{2}_{2}d x^{2}=a(t)r d
\theta, \qquad e^{3}=e^{3}_{3}d x^{3}=a(t)r \sin(\theta) d r.
\end{equation}
Using (\ref{eq forma di connessione}) and the definitions in
(\ref{der spin}) we have (see the Appendix)

\begin{equation}\label{derivata covariante spinoriale componente zero}
  D_{0}=\partial_{0}+\frac{1}{4}[\gamma_{i},\gamma_{0}]\;{\cal H}\;e^{i}_{0}+\frac{1}{4}
  [\gamma_{1},\gamma_{j}]\;\frac{\rho}{a r}\;e^{j}_{0}+\frac{1}{4}
  [\gamma_{2},\gamma_{3}]\;\frac{\tan(\theta)}{a
r}\;e^{3}_{0},\qquad \qquad i=1,2,3;\qquad j=2,3.
\end{equation}

Since we choose diagonal tetrads every term $e^{k}_{0}$ with
$k=1,2,3$ is null. This implies that all the terms but the first
are null whatever the value of the commutators of the Dirac
matrices. It is worth to stress that this result is independent of
the choice of tetrads because the (0,0) component of the energy
momentum tensor of a fluid is equivalent to the energy density
only if the tetrads (or in general the chosen coordinates) are
time-orthogonal as in (\ref{tetradi di Friedmann}). If we choose
non time-orthogonal tetrads, i.e. $e^{0}_{0}$ not constant,
$T_{00}$ does not represent the energy density because it acquires
"pressure components" due to the different orientation of the
tetrad. In our calculation these terms are the second, third and
forth term of the (\ref{derivata covariante spinoriale componente
zero}).

Thus the temporal component of the spinorial derivative in the FRW
metric is just the standard time derivative:
\begin{equation}
D_{0}=\partial_{0}.
\end{equation}
This is not surprising if we consider the symmetries of the metric
element (\ref{metrica di friedmann}). Thus, the (0,0) component of
the stress energy tensor is
\begin{equation}\label{sttress energy tensor esplicito}
  T_{00}=T_{00}^{Flat}.
\end{equation}

This allows us to use $T_{00}^{Flat}$ to compute the cosmological
constant. From Eq.(\ref{tmn}) we thus obtain
\bea\
 T_{00}(x) = \frac{i}{2}:\left({\bar \Psi}_{m}(x)\gamma_{0}
\overleftrightarrow{\partial}_{0} \Psi_{m}(x)\right):\eea where
$:...:$ denotes  the customary normal ordering with respect to the
mass vacuum in the flat space-time.

In terms of the annihilation and creation operators of fields
$\nu_{1}$ and $\nu_{2}$, $T_{00}$ is given by
\bea T^{(i)}_{00}= \sum_{r}\int d^{3}{\bf k}\,
\omega_{k,i}\lf(\al_{{\bf k},i}^{r\dag} \al_{{\bf k},i}^{r}+
\beta_{{\bf -k},i}^{r\dag}\beta_{{\bf -k},i}^{r}\ri), \eea
and we note that $T^{(i)}_{00}$ is time independent.

Next, our task is to compute the expectation value of
$T_{00}^{(i)}$ in the flavor vacuum $|0 {\rangle}_{f}$, which, as
already recalled, is the one relevant to mixing and oscillations.
Thus, the contribution $\lan\rho_{vac}^{mix}\ran$ of the neutrino
mixing to the vacuum energy density is:
 \bea\
 {}_f\lan 0 |\sum_{i} T^{(i)}_{00}(0)|
0\ran_f = \lan\rho_{vac}^{mix}\ran \eta_{00} ~.
 \eea

We observe that within the above  QFT formalism for neutrino
mixing
 we have $ {}_f\lan 0 |T^{(i)}_{00}| 0\ran_f={}_f\lan
0(t) |T^{(i)}_{00}| 0(t)\ran_f$ for any $t$. We then obtain
\bea {}_f\lan 0 |\sum_{i} T^{(i)}_{00}(0)| 0\ran_f =
\sum_{i,r}\int d^{3}{\bf k} \, \omega_{k,i}\lf({}_f\lan 0
|\al_{{\bf k},i}^{r\dag} \al_{{\bf k},i}^{r}| 0\ran_f + {}_f\lan 0
|\beta_{{\bf k},i}^{r\dag}\beta_{{\bf k},i}^{r}| 0\ran_f \ri) .
\eea
Since \cite{BV95}
\bea\label{33} {}_f\lan 0 |\al_{{\bf k},i}^{r\dag} \al_{{\bf
k},i}^{r}| 0\ran_f = {}_f\lan 0 |\beta_{{\bf
k},i}^{r\dag}\beta_{{\bf k},i}^{r}| 0\ran_f\, = \,\sin^{2} \theta
|V_{\bf k}|^{2}, \eea
we get
\bea\label{34} {}_f\lan 0 |\sum_{i} T^{(i)}_{00}(0)| 0\ran_f
=\,8\sin^{2}\theta \int d^{3}{\bf
k}\lf(\omega_{k,1}+\omega_{k,2}\ri) |V_{\bf k}|^{2}
=\lan\rho_{vac}^{mix}\ran \eta_{00}, \eea
i.e.
\bea\label{cc} \lan\rho_{vac}^{mix}\ran = 32 \pi^{2}\sin^{2}\theta
\int_{0}^{K} dk \, k^{2}(\omega_{k,1}+\omega_{k,2}) |V_{\bf
k}|^{2} , \eea
where the cut-off $K$ has been introduced. Eq. (\ref{cc}) is our
result: it shows that the cosmological constant gets a non-zero
contribution induced  from the neutrino mixing. Notice that such a
contribution is indeed zero in the no-mixing limit when the mixing
angle $\theta = 0$ and/or $m_{1} = m_{2}$ (see Eq.(\ref{V})). It
is to be remarked that the contribution is zero also in the limit
of $V_{\bf k} \rightarrow 0$, namely in the limit of the
traditional phenomenological mixing treatment.

It is  interesting to note that, for high momenta, the function
$|V_{\bf k}|^{2}$ produces a drastic decrease in the degree of
divergency of the above integral, in comparison with the case of a
free field. Thus, if for example we chose $K\gg \sqrt{m_1 m_2}$,
we obtain:
\bea\label{cc2} \lan\rho_{vac}^{mix}\ran \propto
\sin^2\theta\,(m_2 -m_1)^2 \, K^2 ,
 \eea
whereas the usual zero-point energy contribution would be going
like $K^4$.

Of course, we are not in a position to make our result independent
on the cut-off choice. However, was not this our goal. What we
have shown is that a nonzero contribution to the value of the
cosmological constant may came from the mixing of the neutrinos.
We have not solved the cosmological constant problem. Although it
might be unsatisfactory from a general theoretical point of view,
we may try to estimate the neutrino mixing contribution by making
our choice for the cut-off. Since we are dealing with neutrino
mixing, at a first trial it might be reasonable to chose the
cut-off proportional to the natural scale we have in the mixing
phenomenon, namely $\textbf{k}^{2}_{0}\simeq m_{1} m_{2}$
 \cite{BV95}.

With such a choice, using $K\sim k_{0}$, $m_{1}=7 \times
10^{-3}eV$, $m_{2}=5 \times 10^{-2}eV$, $k_{0}=10^{-3}eV$ and
 $\sin^{2}\theta\simeq 1$ in Eq. (\ref{cc}), we obtain
\bea \lan\rho_{vac}^{mix}\ran =1.3 \times 10^{-47}GeV^{4}\eea

Using Eq.(\ref{ed}), we have agrement with the upper bound given
in Section I:
\bea \Lambda \sim 10^{-56}cm^{-2} , \eea
Another possible choice is the electroweak scale cut-off: $K\sim
100 GeV$. We then have \bea \lan\rho_{vac}^{mix}\ran =1.5 \times
10^{-15}GeV^{4}\eea and \bea \Lambda \sim 10^{-24}cm^{-2} , \eea
which is, however, beyond the accepted upper bound.

\section{Conclusions}
In this paper we have shown that the neutrino mixing can give rise
to a nonzero contribution to the cosmological constant. We have
shown that this contribution is of a different nature with respect
to that given by the zero-point energy of free fields and we
estimated it by using the cut--off given by the natural scale of
the neutrino mixing phenomenon. The different origin of the mixing
contribution also manifests in the different ultraviolet
divergency order (quadratic rather than quartic, see
Eq.(\ref{cc2})).
 The obtained value is consistent
with the accepted upper bound for the value of $\Lambda$. On the
contrary, by using as a cut--off the one related with the
electroweak scale, the $\Lambda$ value is greater than such upper
bound.

It is worth to stress once more that the origin of the present
contribution is completely different from that of the ordinary
contribution to the vacuum zero energy of a massive spinor field.
As we have shown, the effect we find is not originated from a
radiative correction at some perturbative order \cite{Col-Weinb}.
Our effect is exact at any order. It comes from the property of
QFT of being endowed with infinitely many representations of the
canonical (anti-)commutation relations in the infinite volume
limit. Therefore, the new result we find is that it is the {\it
mixing} phenomenon which provides such a vacuum energy
contribution, and this is so since the field mixing involves
unitary inequivalent representations. Indeed, as
Eqs.(\ref{33})-(\ref{cc}) show the contribution vanishes as
$V_{\bf k} \rightarrow 0$, namely in the quantum mechanical limit
where the representations of the (anti-)commutation relations are
all each other unitarily equivalent. Our result thus discloses a
new possible mechanism contributing to the cosmological constant
value.

As a final consideration, we observe that this effect could also
be exploited in the issue of dark energy without introducing
exotic fields like quintessence. In fact, neutrinos constitute a
cosmic background of unclustered components whose mixing and
oscillations could drive the observed accelerated expansion.

\section*{Acknowledgments}

We thank MURST, INFN, INFM and ESF Program COSLAB for partial
financial support.

\appendix

\section{}

For the sake of completeness, we derive the 0-th component
spinorial derivative used to get the results (\ref{sttress energy
tensor esplicito}). We start from the Cartan Equations (\ref{eq
forma di connessione}) writing the tetrads one-forms in the FRW
metric (\ref{metrica di friedmann}) as in (\ref{tetradi di
Friedmann}).

The exterior derivative of $e^{a}$ is
\begin{eqnarray}\label{differenziale esterno tetradi}
d e^{0}&=& dt\wedge dt=0\\ d e^{1}&=& \frac{\dot{a}}{\rho}\;
dt\wedge d x^{1}= \frac{\dot{a}}{a}\; e^{0} \wedge e^{1}\\ d
e^{2}&=& \dot{a}r\; dt\wedge d \theta +a\; dr\wedge d \theta =
\frac{\dot{a}}{a}\; e^{0} \wedge e^{2}+ \frac{\rho}{a r}\; e^{1}
\wedge e^{2}\\\nonumber d e^{3}&=& \dot{a}r\sin(\theta)\; dt\wedge
d \phi +a\sin(\theta) \; dr\wedge d \phi+ a r \cos(\theta)\;
d\theta\wedge d \phi=\\ && = \frac{\dot{a}}{a}\; e^{0} \wedge
e^{3}+ \frac{\rho}{a r}\; e^{1} \wedge e^{3}+\frac{\tan(\theta)}{a
r}\;e^{2} \wedge e^{3}.
\end{eqnarray}
Since the connection forms are antisymmetric in the tetradic
(latin) indexes, Eq. (\ref{eq forma di connessione}) gives
\begin{eqnarray}\label{espressione delle unoforme di connessione}
\omega^{0}_{0}&=&0\\ \omega^{0}_{i}&=&-\frac{\dot{a}}{a} \;
e^{i}={\cal H}\;e^{i}\qquad i=1,2,3\\
\omega^{1}_{j}&=&\frac{\rho}{a r}\;e^{j}\qquad \qquad \qquad
j=2,3\\ \omega^{2}_{3}&=&\frac{\tan(\theta)}{a r}\;e^{3}\\
\omega^{i}_{i}&=&0.
\end{eqnarray}
Using the definition of spinorial derivative and spinorial
connection given in (\ref{der spin}), we have
\begin{equation}\label{derivata covariante spinoriale}
 D_{\mu}=\partial_{\mu}+\frac{1}{4}[\gamma_{i},\gamma_{0}]\;\omega^{i 0}_{\mu}+
 \frac{1}{4}[\gamma_{1},\gamma_{j}]\;\omega^{1 j}_{\mu}+
 \frac{1}{4}[\gamma_{2},\gamma_{3}]\;\omega^{2 3}_{\mu},
\end{equation}
where we have used the antisymmetry of the commutators. The 0-th
component of Eq.(\ref{derivata covariante spinoriale}) gives
 Eq.(\ref{derivata covariante spinoriale componente
zero}).


\end{document}